\documentclass[12pt,preprint]{aastex}
\shortauthors{Saripalli, Subrahmanyan, \& Udaya Shankar}
\shorttitle{Giant radio galaxy PKS~B1545$-$321}
\begin{document}

\title{Renewed activity in the radio galaxy PKS~B1545$-$321: twin 
edge-brightened beams within diffuse radio lobes}

\author{Lakshmi Saripalli, Ravi Subrahmanyan}
\affil{Australia Telescope National Facility, CSIRO, 
Locked bag 194, Narrabri, NSW 2390, Australia}

\and

\author{N. Udaya Shankar}
\affil{Raman Research Institute, 
Sadashivanagar, Bangalore 560 080, India}

\begin{abstract}
Australia Telescope Compact Array (ATCA) images of the 
giant radio galaxy PKS~B1545$-$321 show a pair of
oppositely directed beams emerging from a radio core 
and ending in bright components that are symmetrically located
on either side. These inner beams are embedded within edge-brightened
outer lobes of lower surface brightness and the bright ends of
the inner beams are well recessed from the ends of the outer lobes.
The inner beams and diffuse surrounding lobes  
share a common central core and radio axis.
We propose that the observed inner beams are double lobes which have 
been created within relic outer lobes as a consequence of a
restarting of the central activity; therefore, PKS~B1545$-$321 is a
rare opportunity for examining the development of
restarted beams within a relic synchrotron plasma cocoon.

The inner double representing the new episode has among the
highest axial ratios found in typical edge-brightened 
radio galaxies. The low radio luminosity of the inner double,
the narrow and constant transverse extent of its
cocoon and the relatively low brightness of the hotspots 
at its ends are consistent with the almost ballistic propagation 
expected for a beam that has a low density contrast and is advancing within a 
relatively light ambient medium. 

\end{abstract}


\section{Introduction}

The morphologies of the lobes of double radio galaxies and the variations in
the spectral index over them have recently been used to infer 
the history of the central engine activity 
\citep{roettiger94,subrahmanyan96,schoenmakers00}. 
Such studies indicate that apart from precession  
and other large-angle changes in the central engine axis,
variations in the beam activity also include interruptions:
beams sometimes stop and restart.
Giant radio galaxies (GRGs; \citet{saripalli86}) have proved to be 
a fertile ground for examples of 
interrupted nuclear activity.  \citet{subrahmanyan96} and 
\citet{schoenmakers00} have
drawn attention to several GRGs where morphological properties 
indicated an interruption in central activity. The evidence is in the form of 
a nested pair of radio sources having the same central radio core. Such sources
are excellent opportunities for studying the development of new beams in an
environment that is different from the usual ambient medium --- the thermal 
X-ray gas halo or the thermal intergalactic medium (IGM) --- of typical radio
galaxies. In this respect, it is highly desirable to not 
only have more examples
of such double-double sources but also sources where the new beams are 
observed to be fully
embedded within the lobes of the older activity. Recently, \citet{saripalli02}
presented a detailed study of such a source where the inner double is
clearly imaged and one side is observed to be fully embedded within older 
lobe plasma. Herein we present the finest example yet of 
such a source, PKS~B1545$-$321, in which the inner 
double is traced all the way from the core to its bright ends and 
where both inner components
are observed to be completely enveloped by outer lobe plasma.

An 843 MHz image of PKS~B1545$-$321 was made by \citet{jones92} using the 
Molonglo observatory synthesis telescope (MOST); 
this showed a continuous bridge of emission over its 
entire length along with multiple peaks along a central ridge. 
An ATCA image
of the source \citep{subrahmanyan96}, which was made at 22 cm 
wavelength and with higher angular resolution,  clearly showed
a pair of bright inner compact components lying along the source axis.
These components were observed to be well recessed
from the ends and symmetrically located on either side of
a central radio core.  The unusual morphology was suggested to be
a result of restarted activity after an interrutption. An optical 
host has been identified
at the location of the radio core; the galaxy 
has a $b_{J}$ magnitude of 17.5 and its redshift 
is $z = 0.1082$ \citep{simpson93}.

In this paper we present our new multi-frequency observations of 
PKS~B1545$-$321 using the ATCA,
which have been made with full polarization and with higher dynamic range and  
resolution as compared to previous images.  
The higher quality images have revealed 
the inner components to be a spectacular pair of beams that 
are traced continuously from the central core and 
terminate at bright heads. These twin beams are fully embedded
within diffuse lobes and their ends are well recessed.
We discuss the restarting phenomenon in this source 
and the nature of the 
interaction between the restarted beams and the surrounding 
older cocoon material.

Herein we adopt a flat cosmology with parameters $\Omega_{\circ} = 0.3$, 
$\Omega_{\Lambda} = 0.7$, and a Hubble 
constant $H_{\circ} = 65$ km s$^{-1}$ Mpc$^{-1}$.  PKS~B1545$-$321 is at
a luminosity distance of 488 Mpc.  
The source has a total angular size of $9\farcm0$ and we derive the
proper linear size to be 1.04 Mpc. 

\section{ATCA radio images of PKS~B1545$-$321}

The radio source was observed using the 
East-West ATCA separately 
in a compact 375 m, an intermediate 1.5 km and an 
extended 6 km configuration.  The observations were aimed at
imaging the largest angular scale structures in this $9\arcmin$
size source with a few arcsec resolution at 22 and 12 cm wavelengths.
Each of the three Earth-rotation
Fourier synthesis observations were made in full polarization and
simultaneously using two frequency bands 128 MHz wide centered at
1384 and 2496 MHz. The continuum bands were covered in 13 independent
channels. The flux scale was set
using observations of PKS B1934$-$638 whose flux density was adopted to
be 14.94 and 11.14 Jy respectively at 22 and 12 cm.
The interferometric visibility data were calibrated and imaged 
using {\sc miriad};  data in both observing bands were
separately reduced. 
Continuum images were constructed from the multi-channel data
using bandwidth synthesis techniques and the images were deconvolved
using the Steer-Dewdney-Ito algorithm.  Three iterations of
phase self calibration were performed.

The total flux density of PKS~B1545$-$321
is 1.8 Jy at 1384 MHz and 1.1 Jy at 2496 MHz based on the new
ATCA data.  In Table 1 we summarize the flux density 
measurements for this source. The overall spectrum is 
curved and the spectral index $\alpha \approx 1.1$ between 843 and 2496 MHz
($S_{\nu} \propto \nu^{-\alpha}$). 

A 12 cm wavelength continuum image, showing the entire source, is
in Fig.~1. The linear gray scale representation shows  (1) oppositely 
directed collimated beams terminating in bright components, 
(2) low surface brightness edge-brightened double lobes which appear
to surround the twin beams and extend well beyond, 
and (3) an unresolved core at the common center.
We refer to the first and second structures as the inner and
outer doubles respectively; the northern and southern lobes of
the inner double are called N2 and S2, those of the outer
double are called N1 and S1.
Both the doubles are collinear with the compact 
central core component. 

\subsection{The Outer Lobes}

The structure in the outer lobes at 22 cm is shown in Fig.~2.  
Transverse to the radio axis, the width of the radio source is
fairly constant over most of its length. Both lobes have a similar overall 
morphology and in both cases the ratio of the lobe length to lobe width is
$\approx 3$. Both lobes 
are edge-brightened and sharply bounded; 
in particular, the NW lobe has a pair of emission peaks along a 
relatively brighter rim at its western end. There are no strong hotspots or 
any compact structures at the ends of the lobes. The two lobes are wavy 
and the side-to-side undulations are inversion symmetric about the core 
position.

In the central regions the bridge flares transverse
to the radio axis. The flaring is different for the two lobes. Whereas 
N1 has extensions on both sides, S1
shows an extension only to one side, the NE side. There is a 
curious ``zone of avoidance'' at the center, that is oriented 
transverse to the radio axis, where the intensity is weak. This gap in 
emission is displaced from the location of the core: whereas S1 stops
short of the core along an edge that is south of the core, 
N1 continues past the core position and stops at an edge south of the core.  
Interestingly, the inner lobes
show no emission gap and S2 appears continuous across the gap observed
between S1 and N1.

The distribution of the spectral index over the lobes, between 12 and 22 cm, 
is shown in Fig.~3.  The 12 cm visibility data were tapered and imaged to produce a 
low resolution image and the 22 cm visibility data were imaged with nominal 
resolution. The deconvolved images, which had similar resolutions, were then 
convolved to a common final resolution of $13\arcsec \times 7\arcsec$.  
Effectively, this analysis ensured that the 
images used for computing the spectral index distribution image shown in Fig.~3
had been constructed from data with similar spatial frequency range.
The observed spectral index is close to 0.6 over the brighter
regions at the NW end of the source; at the SE end the spectral index is 
around 0.7. Overall, there is a steepening of the spectral index from
the ends of the outer lobes towards the center where $\alpha$ 
exceeds 0.9.  In N1 and S1 the regions with the steepest spectra appear to 
avoid the brightest parts; in particular, the radio spectrum is relatively 
steeper with $\alpha \ga 1.0$ along the SW edge of S1. The spectral indices in
these regions, with low surface brightness at 12 cm, are estimated to
have an error of 20\%.

We have determined the distribution of the polarized intensity over the
source at 12 and 22 cm.  The rotation measure (RM) has a mean 
value of $-$14.4 rad m$^{-2}$ over the source with a 1-$\sigma$ scatter of
6 rad m$^{-2}$.  We do not observe any significant changes, within the errors 
in the measurement, in the RM distribution over the source. The distribution in
the 22 cm polarized intensity is shown in Fig.~4; overlaid
are vectors showing the orientation of the projected electric field
with bar lengths proportional to the fractional polarization.
The vector orientations have had the small correction for Faraday
rotation that was derived based on the assumption that the observed RM 
distribution is owing to a foreground screen.
The projected magnetic field, which is perpendicular to the electric field 
orientations shown in the figure, is circumferential along the 
boundaries of the outer lobes. The fractional polarization is enhanced along 
the lobe boundary.  The projected B-field is aligned 
with the total intensity ridges in the two lobes and closely follows the
bends and wiggles along the lengths of the outer lobes; as expected,
the B-field orientation appears to be along the flow in the synchrotron plasma.
However, the distribution in the B-field, in this low-resolution image,
is perpendicular to the source axis over the inner lobe N2 where we expect the
flow in the outer lobe N1 to be along the source axis.  It is likely that
towards N2 the polarized intensity in this image is dominated by emission
from N2 and the observed field represents the orientation in N2 and not 
in N1.  The polarized intensity distribution shows much 
more structure as compared to the total intensity distribution.
The fractional polarization is distinctly lower in the regions
of the outer lobes close to the ends of N2 and S2; these 
can be recognized as being due to beam depolarization because the 
E-vectors here are observed to sharply change orientation. 
There are also regions within the outer lobes where polarized intensity
is low and, consequently, undetected.  They are observed to be located
somewhat offset from the ridges, close to where the 
ridge lines have sharp bends,
and preferentially on the concave side.  The low polarized 
intensity at some of these
locations might also be due to beam depolarization.
The wings transverse to the radio axis at the center of the radio galaxy 
show parallel B-fields, this is particularly noticeable 
along the inner edge of S1.  Additionally, the wings have a relatively 
high (30--60\%) fractional polarization. 

To summarize the polarization properties, overall there 
is order in the magnetic field structure and we recognize 
the following components: (1) the parallel B-field that
closely follows the ridges in the two lobes over several hundreds of 
kiloparsecs, (2)  the parallel B-field along the edges of most of the 
source, and (3) the parallel B-field in the wings on both sides of the
``zone of avoidance''.

We estimated the average depolarization ratio (as a 
ratio of the percentage polarization 
at 22 cm to that at 12 cm) over different parts of the outer lobes: 
the ratio is 
about 0.96 at the SE end and close to 1.0 at the NW end.  
In the central parts the ratio falls to about 0.92. 
Depolarization is most pronounced on the 
SW sides of the inner parts 
of  N1 and S1 where the ratio is as low as 0.83. These are the parts with
the steepest spectral indices and, as discussed in section 3.1, these 
parts are on the side where extended X-ray emission has been detected.

\subsection{The Inner Double}

A high resolution 12 cm image of the inner double is shown in Fig.~5
with a beam of FWHM $5\farcs1 \times 2\farcs5$.  We also show, in Fig. 6,
a 12 cm image made excluding visibilities 
with spatial frequencies less than 
3 k$\lambda$; this image reproduces only structure with angular scales
smaller than about $0\farcm5$ 
and shows the inner lobes without the diffuse underlying outer 
lobe emission.  The total flux density of the inner double is 
53 and 78 mJy respectively at 12 and 22 cm. 
The inner double is observed to be twin oppositely directed beams that are
traced continuously from the core and terminating at bright 
axially elongated cylindrical components.  N2 and S2 consist of bright ends,
short trails, and relatively lower-surface-brightness emission connecting the
ends to the core. The striking feature of the inner double is the near
constancy of its width over its length; this is best represented visually 
in the 12 cm total  intensity gray-scale image in Fig. 1.  The deconvolved 
FWHM, derived using the image in Fig. 6,  is about $6\arcsec$--$7\arcsec$ 
all along the $3\arcmin$ length of the inner double.

The bright ends are resolved in the direction transverse 
to the radio axis and have deconvolved FWHM 
in the range $3\arcsec$--$5\arcsec$ corresponding to a linear size of 
$\approx 8$ kpc;
this is less than the FWHM widths of the short trails and lower surface
brightness parts of the inner double. As shown in Fig.~1, 
these bright peaks at the ends of inner double are the brightest components
in the entire source; however, their surface brightness (at 12 and 22 cm) is 
in the range 0.1--0.4 mJy arcsec$^{-2}$, which is much lower than what is
usually observed in the hotspots of powerful radio galaxies.
The peak at the end of S2 is brighter and has a 
smaller deconvolved size --- both
by factors of 1.4 --- as compared to the peak at the end of N2.  
Additionally, the bright trail behind the peak in S2 is observed to extend
more than twice the distance as compared to that in N2. 
The low surface brightness parts of N2 show intensity
variations with maxima  3--4 times the minima; the S2 beam is 
more uniform. 

The ATCA observations presented in this paper show that the bright 
trails in N2 and S2, i.e., on both sides of the inner double, have double 
peaked transverse profiles.  In Fig. 7, we show a slice profile made 
transverse to the axis in the bright trail close to the end of  S2 to display
this characteristic.  The observations
suggest that the emission is predominantly from a cylindrical sheath. 
In the lower surface brightness parts closer to the core,
the present observations do not have the surface brightness sensitivity 
(at this resolution and observing frequency) to conclusively differentiate 
between edge-brightened and filled emission.  

The 12-cm polarization in the inner double is shown in Figs. 8 \& 9 
with the same 
resolution as the total intensity distribution in Fig. 5.  
The polarized intensity is shown in Fig. 8 using gray scales.  
The electric vectors are 
shown in Fig. 9 with lengths representing the fractional polarization and 
with their orientations corrected for Faraday rotation assuming
a uniform RM of $-$14.4 rad m$^{-2}$.  
At the end of S2, there are two peaks in polarized intensity 
located on either side of the peak in total intensity.  
In the bright trail at the end of S2 there is an enhancement of the 
polarized intensity, as well
as the fractional polarization, along the edges.  Similarly, at 
the end of N2, the peaks in polarized intensity
avoid the total intensity peak and are located along the boundary.  
In the brighter regions at the ends of the inner double the projected 
magnetic field is circumferential along the edges and the fractional
polarization takes on values as high as 45\%.
The distributions of the polarized intensity, as well as that of the
fractional polarization, show an additional ridge 
running along the middle on both sides of the inner double.  
Along this central
line, the magnetic field is oriented perpendicular to the source
axis and the percentage polarization reaches 40\%. 
The variations in the polarized intensity transverse to the axis
are perhaps because of beam depolarization owing to a flip in the 
orientation of the linear polarization in going from the central ridge line 
to the edges.  

The distribution in the spectral index $\alpha$ over the inner lobes
is shown in Fig. 10.  
Images of the inner lobes at 12 and 22 cm were
separately constructed using visibilities restricted to a common range
of 3--29 k$\lambda$, both images were convolved to a common final
beam of FWHM $11\arcsec \times 7\arcsec$ and these were used in deriving
the distribution in the spectral index.  The spectral index is fairly 
constant at about $\alpha \approx 0.7$
over the bright ends and associated trails in the inner double. However,
there is some evidence for a steeper spectral index with $\alpha \ga 1.0$
in the low brightness parts closer to the core.

The inner double is curved over its entire length; N2 and S2 bend to the SW 
with a mirror symmetry.  If we extrapolate the curved 
trajectories of the inner double, N2 meets the end of the N1 at the 
recessed peak along the western rim and S2 
appears heading for the SW tip of S1.

\subsection{The radio core and host galaxy}

The radio core is unresolved in our images and has a 
deconvolved size $< 1\arcsec$ FWHM.  
The core flux density is 4.5 mJy at 22 cm
and has a spectral index $\alpha = 0.44$ between 22 and 12 cm.
The core power is
$1.2 \times 10^{23}$ W Hz$^{-1}$ at 1.4 GHz and is 0.25\% of the
total source power.

The host galaxy, coincident with the radio core, 
is derived to have an absolute magnitude $M_{B}=-20.9$.
The SuperCOSMOS (SCOS) digitization of the 
UK Schmidt plates in the red and blue (Fig. 11) show that
the parent galaxy is symmetric in 
the red but has isophotal distortions and patchiness in the blue. There 
is an indication of a central dust lane oriented orthogonal to the axis
of the radio galaxy.

\section{Discussion}

\subsection{The relic outer lobes}

The source has a total radio power of $5 \times 10^{25}$ W Hz$^{-1}$ at 
1.4 GHz. Given the absolute optical magnitude of the host galaxy, 
the radio power
places the source above the FR I-FR II dividing line \citep{owen93} 
and in the powerful radio galaxy regime. This is consistent with the edge
brightened morphology of the outer lobes and the spectral 
steepening in the outer lobes towards the core. 
However, the outer lobes do not appear to have
any hotspots at their ends. This suggests that the outer
double is not currently being fed by beams from the center and
is a relic of a powerful radio galaxy created by past activity in the 
central engine.  The relatively low
and uniform surface brightness distribution across the outer lobes
and the overall steep spectrum support the relic hypothesis.  

Radio galaxies that are observed to have small 
axial ratios and fat double type
structure tend to have relatively steep radio spectra as compared to other
powerful double radio sources and they are presumed to be 
relics \citep{muxlow91}; 
however,  PKS~B1545$-$321 is not a fat double. The circumferential magnetic 
fields, 
increased percentage polarization towards the lobe edges and the sharp 
boundaries of the outer lobes indicate 
a compression of the lobe plasma as it expands against the external medium.
Similarly, the high fractional polarization
and parallel B-field in the central flare 
suggest a compression of backflowing lobe against material in
the ``zone of avoidance''.  These observations, which indicate that the
lobes are dynamically evolving, may be reconciled with
the lack of hotspots at the ends if the
beams feeding the outer lobes have been switched off relatively
recently.  We find that the core power and the total power 
in PKS~B1545$-$321 are consistent with the correlation found 
between these quantities for active radio 
galaxies \citep{giovannini88}; this again implies that the outer lobes have 
not faded significantly since the beams to 
these lobes were switched off. Additionally,
the concordance suggests that the core power has not changed significantly as
a consequence of the restarting of the beams and that the current beam power 
is likely to be the same as it was prior to the interruption. 
Assuming that the core power was about $1.2 \times 10^{23}$~W~Hz$^{-1}$ at
1.4 GHz while the beams fed the outer lobes, the correlation 
found by \citet{giovannini88} indicates that the total power of the source 
was at most about $10^{27}$~W~Hz$^{-1}$ at 0.4 GHz.  This implies
that the radio power of the outer lobes has reduced by at most a factor
of 10, and - assuming a tangled magnetic field - adiabatically expanded by 
at most a factor of  1.4,
since the beams were switched off.  It may be noted here that the
closeness of the brightness peaks in the outer lobes to the boundaries
of the lobes indicates that the true expansion factor is much less than
this limit.  If the density of the ambient IGM is 
assumed to be at most 10$\Omega_{B} = 
10 \times 0.019h^{-2}$ \citep{burles99}, the relic outer lobes, which have a 
minimum energy density of 
$\approx 10^{-12}$~ergs~cm$^{-3}$, would expand with a speed of at least
$0.01c$ assuming ram pressure confinement.  Together with the limit on
the expansion factor, we infer that the time elapsed 
since the beams ceased feeding hotspots at the ends of N1 and S1
is $< 7 \times 10^{7}$ yr.   

N1 is observed on the sky to extend beyond the core whereas S1 appears 
to be docked  100 kpc short of the core.  In the lobes of powerful double 
radio 
galaxies that extend to the center, the
bridges are either continuous or are observed to be 
docked symmetrically about the core.  
Assuming that lobes do not backflow past the core, 
the observed structure in PKS~B1545$-$321 
suggests that the source is not on the plane of the sky 
and, additionally, the lobe structure is 
intrinsically asymmetric about the core.  N1 and S1 are 
docked at different distances 
from the core; additionally, the deviations from axi-symmetry 
may be different for the two lobes.
 
We have extracted archival ROSAT total-band PSPC images of the field,
smoothed the $15\arcsec$-pixel counts with a Gaussian of FWHM $1\arcmin$,
and displayed the X-ray brightness distribution in Fig. 12.
Contours of the 22-cm radio intensity are overlaid.
Extended low surface brightness X-ray emission is observed with a peak 
at a location offset to the SW of the radio core. 
The sky positions of objects, which are classified as galaxies (based on
SCOS digitization of the UK Schmidt red and blue plates) 
and are cataloged in SCOS to 
have magnitudes within $\pm3$ of that of the host galaxy, are 
marked in the figure.
The GRG is in a neighbourhood with a high galaxy count: this has 
been noted by \citet{subrahmanyan96} who found that in the vicinity of 
this GRG the sky density of galaxies within $\pm 2$ mag. 
of the host galaxy is a factor of 6 greater than the mean.
We have examined the galaxy distribution within different magnitude ranges
centered at the magnitude of the host galaxy and find
no evidence for any concentration of galaxies at the location of the 
X-ray peak.  

The X-ray gas is not observed to extend along the ``zone of avoidance''
where there is a deficit of synchrotron emission.
However, the central radio wing to the SW of the core
is observed to extend towards the X-ray peak: this radio component occupies
a gap in the X-ray emission and the X-ray contours follow the radio
contours suggesting that the synchrotron 
plume has pushed aside the thermal gas.  The close correspondence
between the X-ray and radio emission suggests a physical interaction 
between the radio plume and the gas responsible for the X-ray emission.  
X-ray gas with comparable brightness is not observed on 
other sides of the radio source and this argues against a
scenario where the X-ray gas was centered on the host galaxy before it was
displaced by the synchrotron gas. 

\citet{strom88} suggested that thermal
gaseous halos, which are ubiquitous in giant ellipticals, 
may be the cause for gradients in the observed
Faraday depolarization along radio lobes. In the case of 
PKS~B1545$-$321, the offset thermal gas environment observed
in the PSPC images, along with any
thermal gas halo associated with the host elliptical galaxy,
may be responsible for the increased depolarization 
observed towards the central regions of the outer lobes and, in particular, 
towards the SW parts.  

\subsection{The renewed activity}

Although the outer lobes appear to be relics of a powerful radio
source in which the energy injection via beams has recently ceased,
the nucleus is currently active as evidenced by the visibility of the 
radio core.  
The current activity appears as twin beams with bright ends
that are well recessed from the ends of the source.  At these ends
the B-field is circumferential and the polarized intensity peaks
are on either side of the total intensity peaks suggesting that 
the beams terminate at these relatively bright components 
located well short of the ends of the outer lobes. When powerful beams that 
terminate in hotspots at the ends of FR II 
double radio sources decline in beam power, we would expect the forward 
advance speeds of the hotspots to decline and the decreased beam 
power may also impact on the beam collimation and stability; we would 
not expect the hotspots to recess or retreat.
The two double radio structures share a common core
component and a common radio axis, both are likely 
to have been produced as a result of different episodes of 
nuclear activity in this common central engine and the restarting
has happened without any change in the direction of the ejection axis.  
PKS~B1545$-$321 appears to be a striking example of restarted activity
in a powerful radio galaxy in which new beams are observed in the
act of ploughing through the relict cocoon of past activity.

The 1.4 GHz total power of the inner double is $2 \times 10^{24}$ W~Hz$^{-1}$ 
placing it below the FR I-FR II divide.  
The inner double has a linear size of 340 kpc and if we compare with
3CRR sources in the size range 300--400 kpc, the inner 
double in PKS~B1545$-$321 has a radio
power that is more than an order of magnitude below the lowest power
observed for edge brightened sources.
Given its relatively low power,
the structure of the inner double might be expected to 
resemble edge-darkened sources and have bright inner jets and a bright core.
However, the morphology of the 
inner double is edge brightened and this is typical of 
powerful FR II radio sources spanning a wide range of linear sizes. 
Additionally, no bright jets are observed and the core 
contains less than 5\% of the 
total power of the inner double: these properties are also 
more representative of FR II type sources \citep{muxlow91}.   

The inner double has an FR II morphology. The relation between the optical 
magnitude of the host galaxy versus radio power \citep{owen93} indicates 
that an FR II type double source, with an optical magnitude corresponding 
to the host of PKS~B1545$-$321, would have a 1.4 GHz radio power in the range 
$8\times 10^{24}$--$2\times 10^{27}$~W~Hz$^{-1}$
if it were evolving in an IGM environment.  The observed radio power of the
inner lobe is lower by a factor in the range 4--1000.   FR II sources
in IGM environments are inferred to be advancing with speeds a few percent 
of the speed of light \citep{scheuer95}.  If we assume that the low observed
power in the inner lobes is owing to a 1D expansion of the  
post-hotspot material (which has a tangled B-field) along the source axis and 
in the relatively low density cocoon environment, we infer
that the inner lobes are advancing with speeds that are a factor 2--64 
greater than that in 
sources in IGM environments.   The inner double in PKS~B1545$-$321 is 
expanding with speeds exceeding 0.1$c$ and its age is inferred to be 
less than $< 5 \times 10^{6}$ yr.

Edge-brightened radio galaxies over a wide range of linear sizes from
parsecs to several megaparsecs have overall a self-similar 
morphology \citep{alexander00,subrahmanyan96}; the inner double
in PKS~B1545$-$321, although it is edge brightened has a very different 
structure.
The lobes of the inner double are about 12--14 kpc wide all along their 
lengths. This width is larger than known jet widths; however, the high axial
ratio lobes do not resemble the cocoons that we see 
typically associated with FR II radio galaxies. The inner double, viewed as a
separate radio source,  has one of the highest 
axial ratio lobes among known edge brightened radio galaxies 
and, as mentioned in 
section 2.2, a remarkable feature of the inner double is its almost 
constant width. The high axial ratio in which the total length 
is about 26 times the width,
is nearly twice larger than the highest axial ratio observed 
among 3CRR radio galaxies that have sizes 300--400 kpc. 

The indications are that the unusual
circumstances in which the new beams are evolving --- emerging into a relic
synchrotron plasma instead of ambient thermal gas --- dramatically alters the
dynamical evolution resulting in an edge-brightened structure, with a 
high axial ratio, in a relatively low luminosity source.  

\subsection{The structure and development of the restarted beams}

The edge-brightening observed in total intensity 
along the boundaries of the short trails at the ends of N2 and S2, together
with the flip in the B-field from being transverse to the radio axis along the
centre to a parallel configuration at both boundaries, suggests that the 
flow is along the center and post-hotspot plasma forms a sheath or cocoon.
The enhanced fractional polarization at the boundaries, together with the
parallel B-field configuration, suggest a compression of this sheath plasma.
Although the inner lobes are observed to be of constant width, implying
that the lateral expansion is insignificant, the surface brightness in
the inner lobes decreases and the spectral index steepens towards the core. 
The inner lobe properties, including the width and surface 
brightness, are observed to be continuous and unchanged as 
S2 traverses the ``zone of avoidance''.

The new beam propagates into the cocoon formed in the past activity phase. 
The ambient medium is, therefore, not galactic or intergalactic thermal gas,
as is usually the case, but synchrotron plasma. The two parameters that
govern the beam propagation: the Mach number of the jet flow, which depends
on the sound speed in the ambient gas, and the ratio 
of the density of the jet material to the ambient medium will both be unusual.
The density of the synchrotron plasma is expected to be
lower than that for an IGM environment unless there is substantial
entrainment. The lobes of GRGs are overpressured
with respect to the IGM \citep{subrahmanyan93,saripalli02}.
In cluster environments, where the ambient intra-cluster medium (ICM) 
surrounding radio sources has been observed,
the radio lobes create synchrotron cavities
sweeping aside most of the ambient gas \citep{boehringer93}.  
These suggest that the sound speed
in the environment of the inner lobes is higher as compared to the IGM
environment encountered at the ends of the beams that created the outer lobes.
Additionally, the density contrast between the beams and the environment is 
lower for the restarted beams.
Cocoon formation is pronounced when light, supersonic beams encounter 
high densities \citep{norman82}; in the case of the restarted beams 
only weak shocks are expected as the (mildly supersonic) beam impinges on 
the relic synchrotron cocoon. The beam is expected to propagate 
almost ballistically and the formation of strong hotspots as well as 
cocoons is inhibited. The shocked post-hotspot jet 
material would not spread out; instead, it is expected to move forwards 
trailing behind the hotspot \citep{williams91}. 
The extremely narrow, constant width and weak cocoon with
decreasing surface brightness towards the center
that we see in the inner 
double of PKS~B1545$-$321 could be representing just such a trail. The surface 
brightness at the ends of the inner double is significantly lower than the
typical surface brightness of hotspots in edge-brightened radio galaxies 
\citep {hardcastle98}: this is consistent with the weak Mach disks 
that were produced in the simulations of  restarted jets 
made by \citet{clarke91}. 
On the other hand, we do not detect any 
feature in the surrounding outer lobes resembling the bow shock predicted 
in their simulations.  

The forward advance speed of the inner lobes is expected to be higher than
that in powerful sources in which beams encounter relatively denser IGM or 
inter-stellar medium \citep{clarke91}.
The exceptionally high axial ratio of the inner double as compared to sources 
of similar size suggest an unusually high axial advance speed for the inner 
double. In addition, the parallel B-field configuration seen in the cocoon 
can be representing the natural outcome of anisotropic expansion of a tangled 
field \citep{leahy91}. As discussed in the section 3.2, the greater expansion 
losses could be the cause for the relatively low luminosity of the inner 
double as compared to edge-brightened sources that have similar sizes.
The axial expansion of the cocoon plasma, due to forward momentum of the 
post-hotspot material, will have an effect on the spectral index distribution 
in the case of a curved electron energy spectrum and may be a 
contributing factor for the spectral steepening seen away from the ends. Ageing
of the synchrotron plasma after it leaves the hotspot is commonly held as the
main cause for the gradual steepening of the spectral index towards the core; 
however, the inner double has been inferred to have an age 
$<5 \times 10^{6}$ yr and significant spectral steepening is not expected 
in this time.  

A few rare examples of radio galaxies, such as the GRG 4C39.04, 
do have constant width and high axial ratio lobes 
similar to that in the inner double. Constant width cocoons in these sources
have been attributed to static pressure balance (as suggested for 4C39.04 
by \citet{hine79}) or to ram pressure balance in the 
transverse direction against a  
falling external density medium (as suggested for the GRG 8C 0821+695 by 
\citet{lacy93}). The inner double in PKS~B1545$-$321 is developing within
the outer relic lobes that have displaced most of the intergalactic 
thermal material and the uniformity in the brightness of the outer
lobes in the vicinity of the inner double implies that here the 
synchrotron plasma,
including any thermal entrainment, is uniform.  Therefore,
ram pressure in a falling external density is unlikely to be 
the cause for the constant width of the inner double.

The inner lobes as well as their
environment --- the relic outer lobes --- are both synchrotron plasma
and, unless entrainment is significant or there are departures from minimum
energy conditions, differences in surface brightness reflect differences 
in internal pressures.
If we assume minimum energy conditions, the short trails behind the
two ends of the inner double are 10 times over-pressured 
with respect to the surrounding outer lobe plasma. Regions of lower
surface brightness farther from the ends and towards the core are also
over-pressured, but by smaller factors.  The limb-brightening in total
intensity and fractional polarization, together with the parallel B-fields
along the boundaries of the inner lobes, support the 
suggestion that the inner lobes are over-pressured and 
not statically pressure confined.  
In this context, the constancy of the width of the inner lobe suggests
that either the lateral expansion speed (which results in a ram pressure
confinement of the inner lobes) is much smaller than the forward 
advance speed of the ends of the inner lobes, or 
that the inner lobes are confined by
other mechanisms (e.g. magnetic confinement).

On both sides, the inner double is observed to gradually 
bend over its entire length.  The bending is reflection symmetric about the 
core.  The observations imply that the jet is bent and that it follows a curved
path.  The cocoon/sheath follows the jet along its
curved path forming a trail along the bent jet.  Sharp
bends and wiggles that are reflection symmetric might be caused
by accelerations experienced by the host as a result of interactions with
close companions \citep{begelman84}; however, 
the observed bending in the inner double is gradual and all along the length
suggesting that the bending is not due to any momentary interaction.
Such gradual reflection symmetric bends are more likely to be caused by
relative motions of the host galaxy and the surrounding medium; 
however, it is unlikely that any relative
motion between the host galaxy and the IGM will affect
the new beams that are shielded from the IGM by the 
synchrotron plasma of the outer lobes.
The jet probably deviates from a straight path because there is 
a continuous bending arising from an interaction
at the ends of the beams with the surrounding cocoon material.
Large scale density or pressure gradients in the surrounding outer relic 
lobes with a component perpendicular to the radio axis could cause 
bending of both jets towards a common direction.
In such a case, the lateral bending force would have to result in a velocity 
component, transverse to the radio axis, that is about 4\% of the 
forward velocity. The immediate environment
of the jets is the synchrotron plasma in the backflowing outer
lobe and here we do not observe increased 
emissivity (and synchrotron pressure) on the convex side of the bend.
Neither do we observe any 
gradients in RM transverse to the radio axis and in the vicinity of the 
inner lobes. Although there is evidence for a transverse
gradient in depolarization (see section 2.1) it is in the opposite sense to 
that required to account for the bend.

The inner and outer lobes are collinear implying that the ejection
axis has not changed in the restarting.  Unless the channels in which the
beams feeding the outer lobes closed, the restarted beams may be expected
to follow the old paths and, if the past beams followed curved trajectories
just before they stopped we may expect the new beams to follow the same
curved trajectories.  The observation that the trajectories of the 
new beams are towards emission peaks at the ends of N1 and S1
supports such a scenario.

\section{A comparison with restarting inner doubles proposed previously}

Although several cases of radio galaxies with evidence for recurrence 
have been recognized, detailed observations of the inner double structures 
exist in the literature for only
a few: J0116$-$473 \citep{saripalli02}, 3C219 \citep{clarke92}, 
3C288 \citep{bridle89} 
and 3C424 \citep{black92}. 3C219 and 3C424 have a prominent 
stunted jet only on one side.  Inner doubles that represent 
a restarting phase may be 
recognized by two main properties: a
higher surface brightness and a circumferential B-field configuration at 
their ends. The inner double in PKS~B1545$-$321 displays both of these 
properties most clearly. 
This is also the case for the inner double in J0116$-$473.  
In all examples of restarting inner doubles observed to date the inner double
is very narrow and although there is a brightening 
at the ends (weak hotspot)
there are no lobes of the kind observed in normal edge brightened radio 
sources. In 3C424, J0116$-$473 and PKS~B1545$-$321 the cocoon 
is recognized by a trail of emission behind 
the hotspot and this trail has a falling surface brightness towards the core. 
These trails are observed to have sharp
edges indicating that we might be observing a contact discontinuity between
the new cocoon and the relic cocoon. The new cocoons are also observed to 
have projected magnetic fields that are parallel
to the edges: this is further evidence suggesting that the material 
spewn out in the new phase is confined by the relic cocoon environment and that
the inner doubles might not have extensive cocoons beyond what is observed. 
Our observations suggest a constant width for the cocoon 
of the inner double in PKS~B1545$-$321. It is interesting that in the 
re-started source 3C219 as well
there is evidence for a 3--4 times wider cocoon of constant width 
surrounding the jet. 

\section{Summary}

We have presented observations of the giant radio galaxy 
PKS~B1545$-$321 that is a spectacular example
of a source undergoing a renewal of nuclear activity. 
The new activity is manifest as twin, narrow, edge brightened 
radio structures that are symmetric about a radio core
and fully embedded within more extended and diffuse outer lobes. 
The inner double structure has an exceptionally large axial ratio. 
It is suggested that the high axial ratio and absence of wide 
lobes are a result of the evolution in a synchrotron plasma
environment.  The unusual development is also expected to result in
an unusually high forward advance for the ends of the new beams and 
this is probably the cause for the low observed radio luminosity
for the inner double.
Improved imaging of the inner lobes with the VLA are underway and
these will be presented along with a model for the evolution
of the inner double in a later paper.

The relic outer lobes in this source have sharp edges and an 
axial ratio similar to the typical values for powerful radio galaxies. 
These outer lobes are also characterized by a close correspondence
between the total intensity structures and magnetic field all along 
the large scale structure.  The dynamically evolving state of the
outer lobes is interpreted as implying that the beams to the outer
lobes switched off relatively recently and that the restarting has occured
on a timescale small as compared to the dynamical timescale of the
outer lobes.

\acknowledgments

The Australia Telescope Compact Array (ATCA) is part of the Australia 
Telescope, which is funded by the Commonwealth of Australia for
operation as a National Facility managed by CSIRO. 
This research has made use of data obtained through the High Energy 
Astrophysics Science Archive Research Center Online
Service, provided by the NASA/Goddard Space Flight Center. We also 
acknowledge the use of SuperCOSMOS, an advanced photographic plate
digitizing machine at the Royal Observatory of Edinburgh, in obtaining the
digitized image of PKS B1545$-$321 presented in the paper.

\clearpage

\begin{deluxetable}{lcl}
\tablewidth{0pc}
\tablecolumns{3}
\tablecaption{Radio flux density measurements of  PKS B1545$-$321. \label{tab1}}
\tablehead{
\colhead{Frequency (MHz)} & \colhead{Flux density (Jy)} & \colhead{Reference}}
\startdata
     80   &   14    &   \citet{slee77}  \\
    843   &    3.6  &   \citet{subrahmanyan96}  \\
   1384   &    1.8  &   This work  \\
   2496   &    1.1  &   This work  \\
   5000   &    0.29 &   \citet{shimmins74}  \\
   8400   &    0.07 &   \citet{wright91}  \\
\enddata
\end{deluxetable}

\clearpage

\begin{figure}
\epsscale{0.9}
\plotone{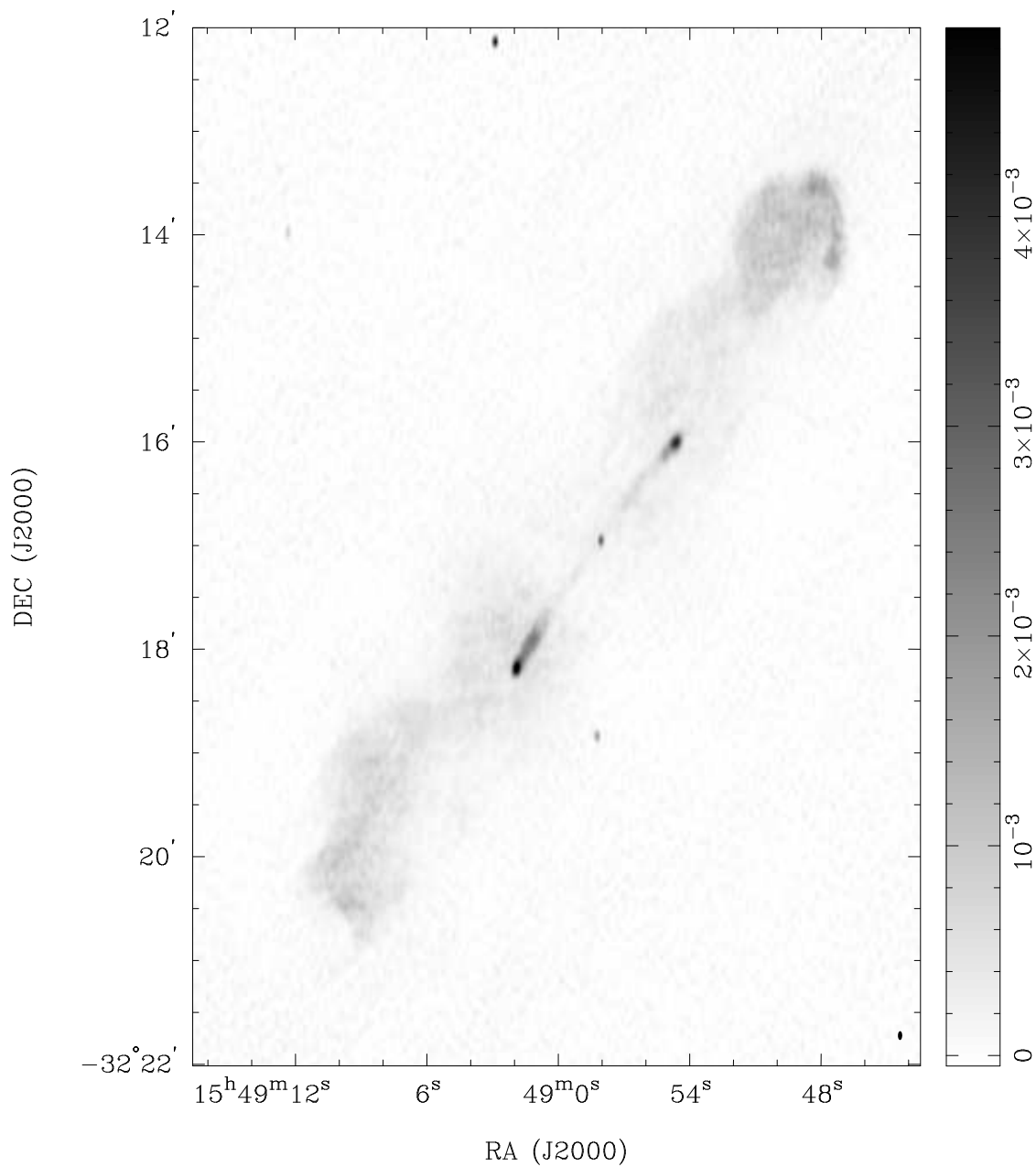}
\caption{Shows the 12~cm image of PKS~B1545$-$321
made with a beam of FWHM $5\farcs1 \times 2\farcs5$  at a 
P.A. of $-0\fdg8$.  Gray scales are shown in the 
range $-0.05$--5~mJy~beam$^{-1}$ with a linear scale. The rms noise
in the image is 50~$\mu$Jy~beam$^{-1}$.
This image, as well as all others displayed 
herein, has been corrected for the attenuation due to the primary beam;
the shaded ellipses in the lower right corner of the images show the
half-power size of the synthesized beams. \label{fig1}
	}
\end{figure}

\clearpage

\begin{figure}
\epsscale{1.0}
\plotone{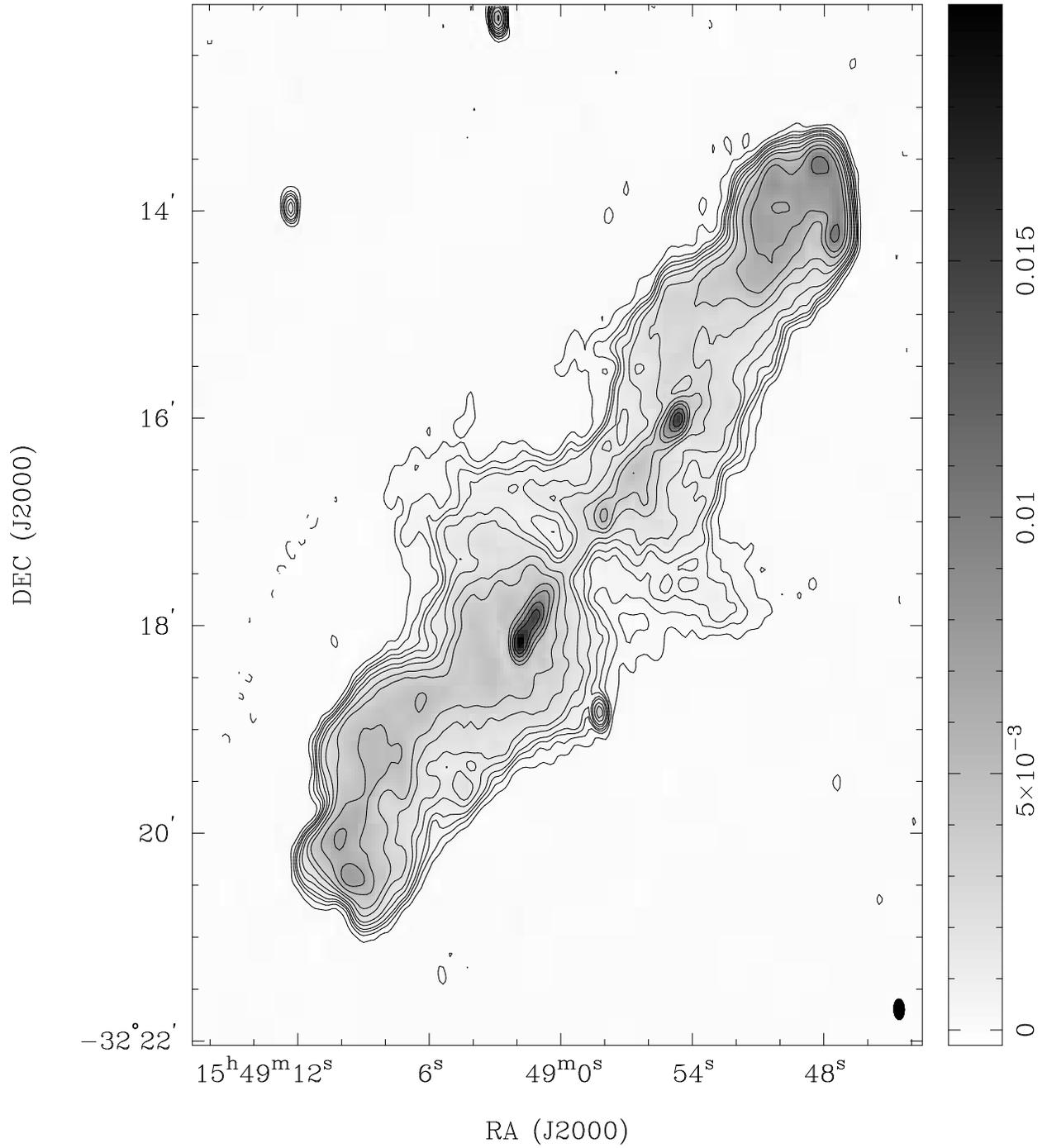}
\caption{Shows the 
22~cm image of PKS~B1545$-$321
made with a beam of FWHM $11\farcs6 \times 5\farcs7$  at a 
P.A. of $1\fdg6$. Contours are at 0.2 mJy~beam$^{-1}$ x ($-1$, 1, 2,
3, 4, 6, 8, 12, 16, 24, 32, 48, 64 and 96). 
Gray scales span the range $-0.3$--20~mJy~beam$^{-1}$. 
The rms noise in the image is 65~$\mu$Jy~beam$^{-1}$.
\label{fig2}
	}
\end{figure}

\clearpage

\begin{figure}
\epsscale{0.9}
\plotone{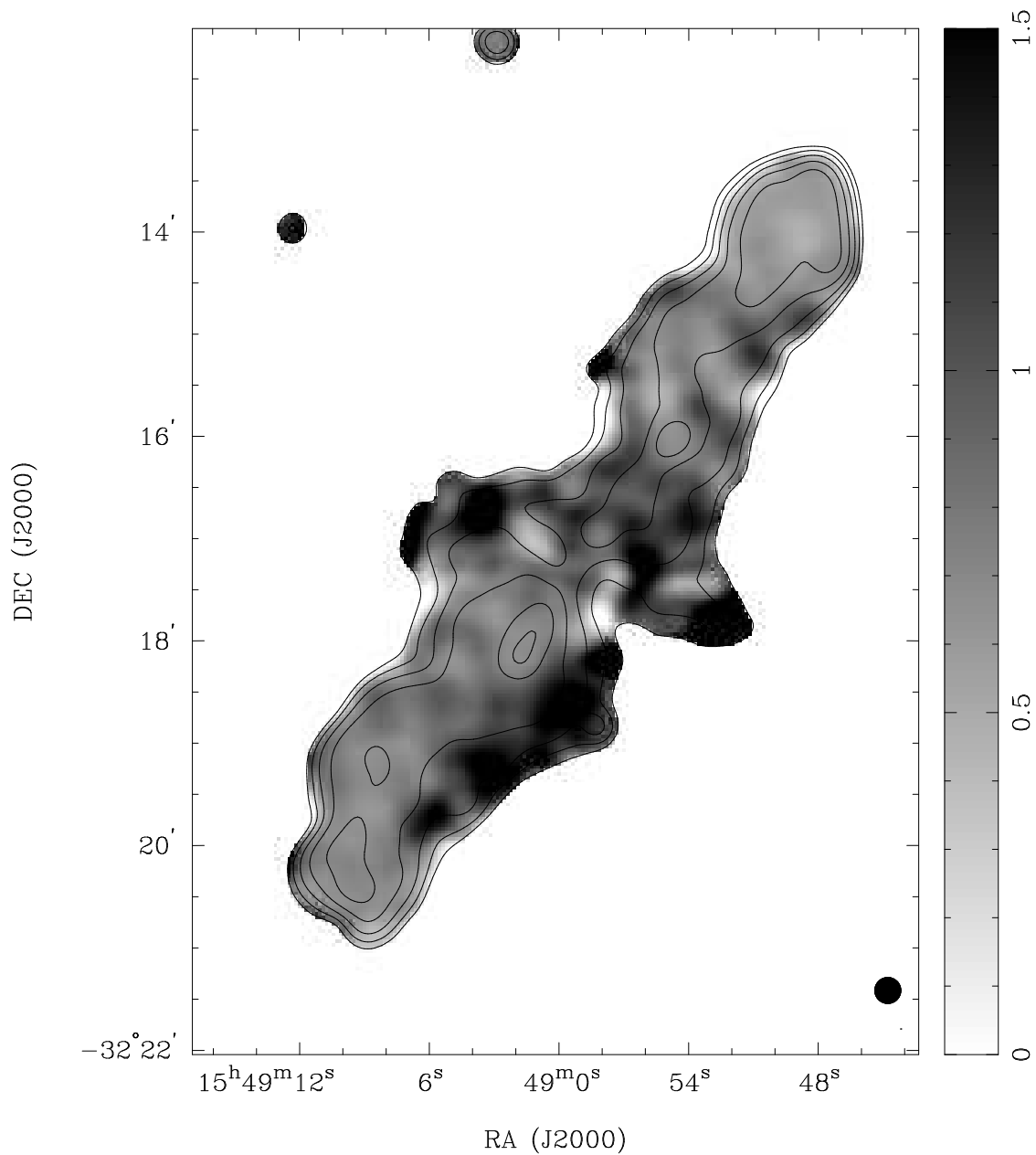}
\caption{Spectral index distribution over the outer lobes of PKS~B1545$-$321. 
The spectral index $\alpha$ ($S_{\nu} \propto \nu^{-\alpha}$) 
has been computed
from images at 12 and 22~cm made with beams of $15\arcsec$ FWHM.  The 
spectral index is shown using gray scales in the range 0--1.5 with the
22~cm contours of total intensity at 1, 2, 4, 8, 16 and 32 mJy~beam$^{-1}$ 
overlaid. \label{fig3}
	}
\end{figure}

\clearpage

\begin{figure}
\epsscale{0.85}
\plotone{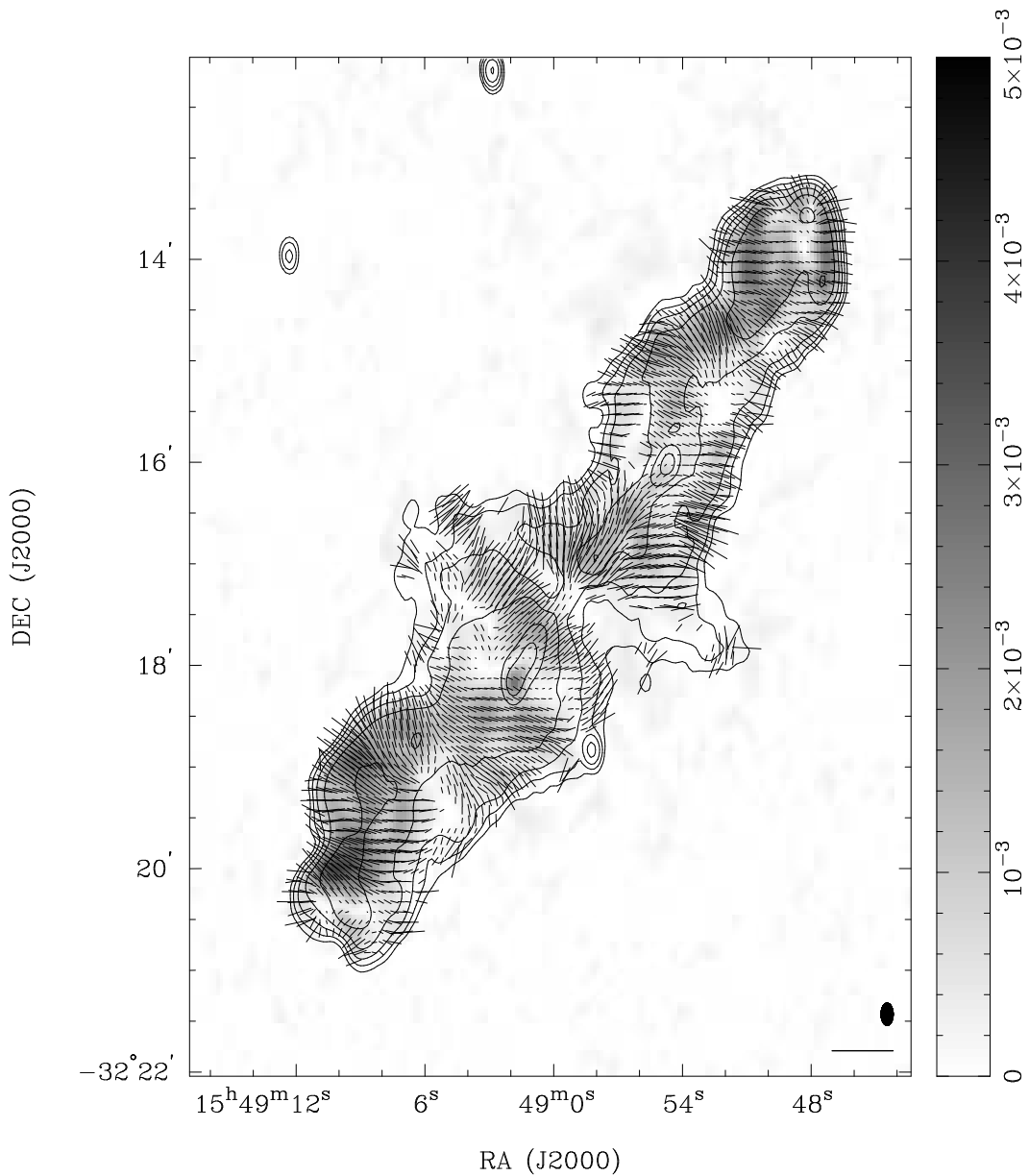}
\caption{The distribution of linear polarization over the source
as observed with a beam of FWHM $13\arcsec \times 7\arcsec$  at a 
P.A. of $0\degr$.  Electric field vectors are displayed with
lengths proportional to the fractional polarization at 22 cm; the
vector shown at the bottom right corner corresponds to 100\%
polarization, the 
orientations of the E-vectors have been corrected for the line of
sight Faraday rotation.  The 22-cm polarized intensity
distribution is shown using gray scales over the range 0--5 mJy~beam$^{-1}$.
Contours of 22-cm total intensity are overlaid; contours are at
0.4 mJy~beam$^{-1}$ x (1, 2, 4, 8, 16 and 32). \label{fig4}
	}
\end{figure}

\clearpage

\begin{figure}
\epsscale{1.0}
\plotone{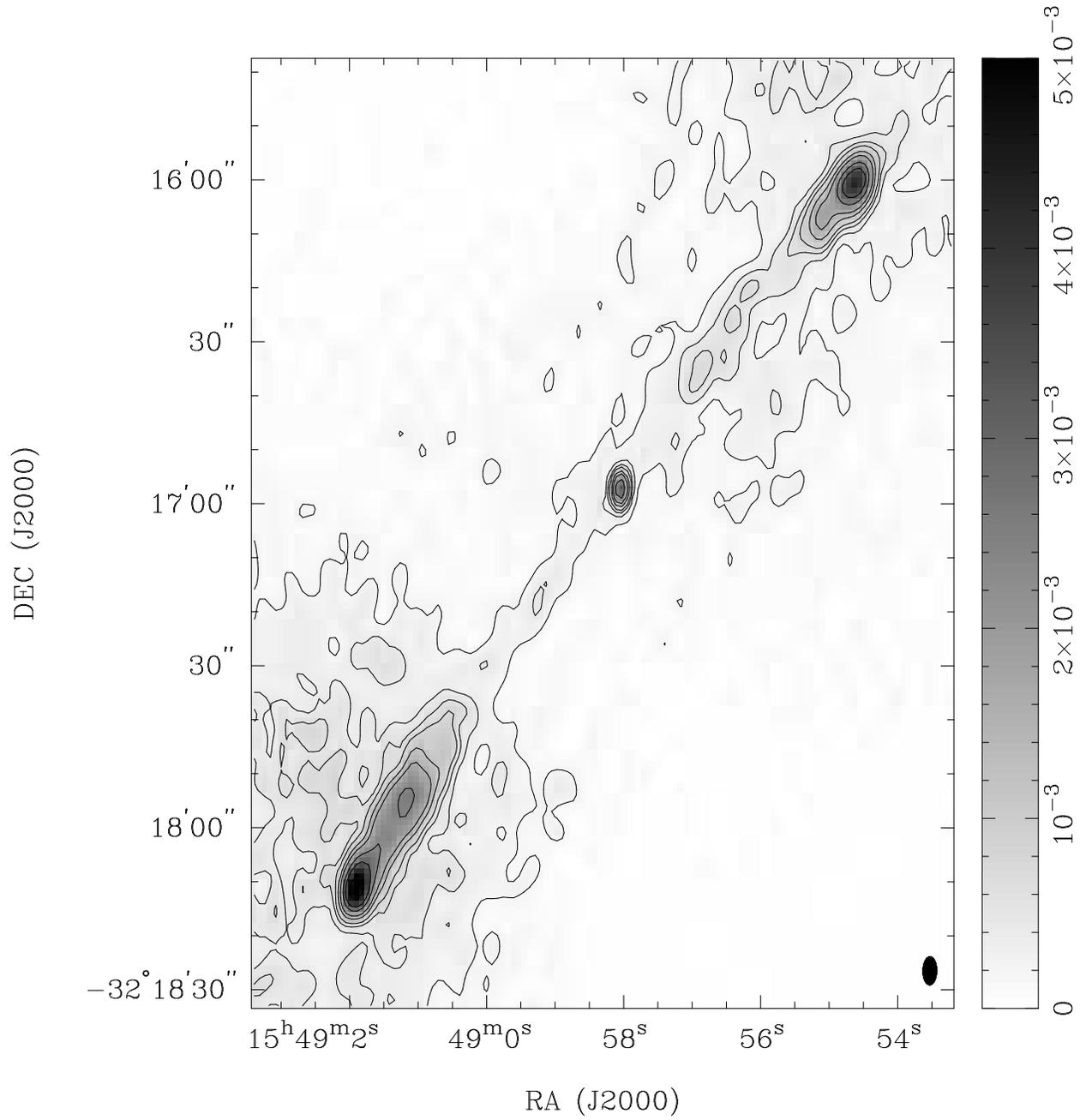}
\caption{Shows the 12~cm total intensity image of the inner lobes made with a 
beam of FWHM $5\farcs1 \times 2\farcs5$ at a 
P.A. of $-0\fdg8$.  Gray scales are shown in the 
range $0$--5~mJy~beam$^{-1}$ with a linear scale. Contours are at
0.2 mJy~beam$^{-1}$ x ($-1$, 1, 2, 3, 
4, 6, 8, 12, 16, 24, 32 and 48). \label{fig5}
}
\end{figure}

\clearpage

\begin{figure}
\epsscale{1.0}
\plotone{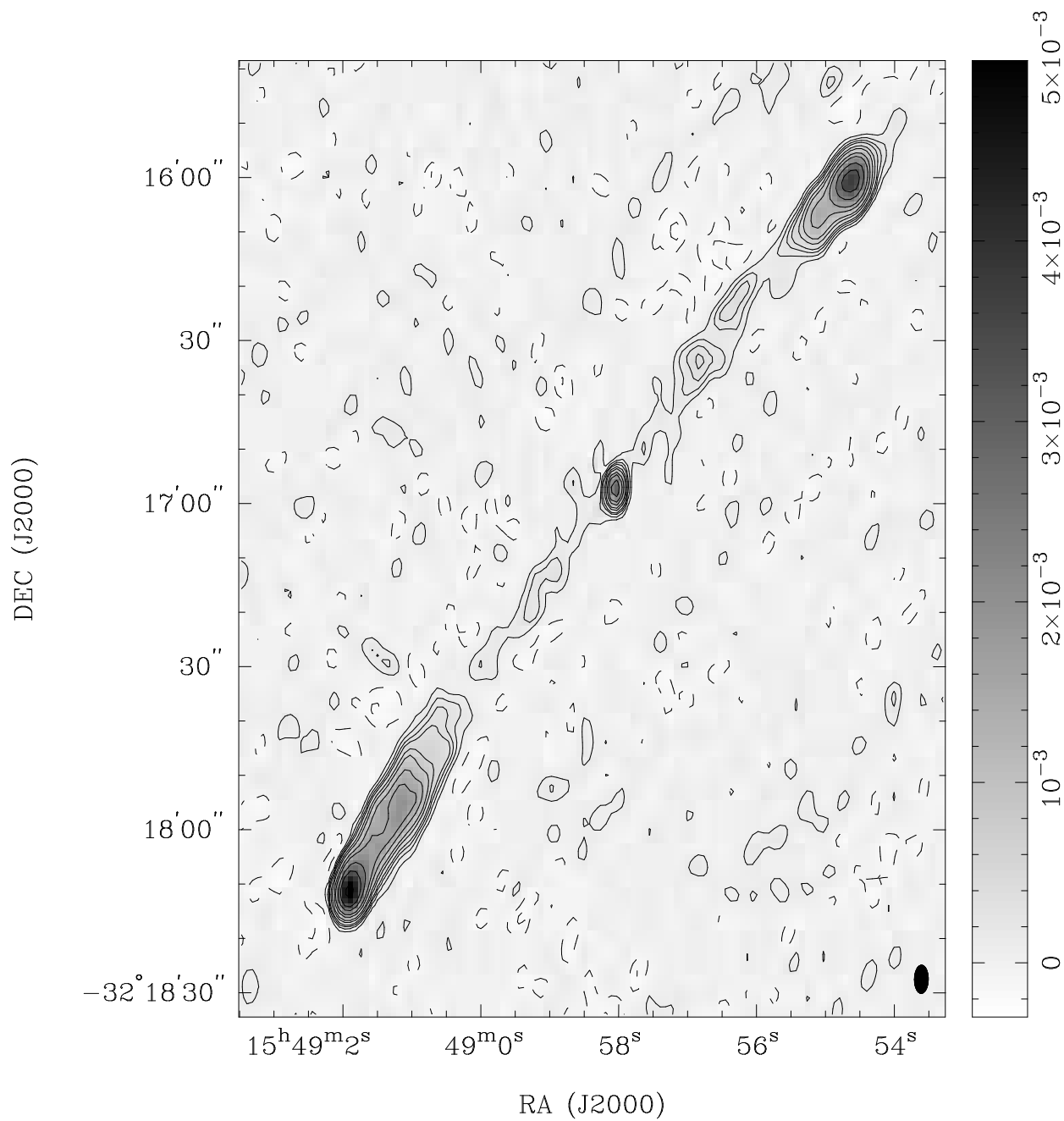}
\caption{
Shows the 12~cm total intensity image of the inner lobes of PKS~B1545$-$321
made using only visibilities exceeding 3 k$\lambda$.
The beam FWHM is $5\farcs1 \times 2\farcs4$  at a P.A. of
$-0\fdg8$.  Contours are 
at 0.1 mJy~beam$^{-1}$ x ($-$2, $-1$, 1, 2, 3, 
4, 6, 8, 12, 16, 24, 32 and 48). Gray scales are shown in the 
range $-0.3$--5~mJy~beam$^{-1}$. \label{fig6}
	}
\end{figure}

\clearpage

\begin{figure}
\epsscale{0.8}
\plotone{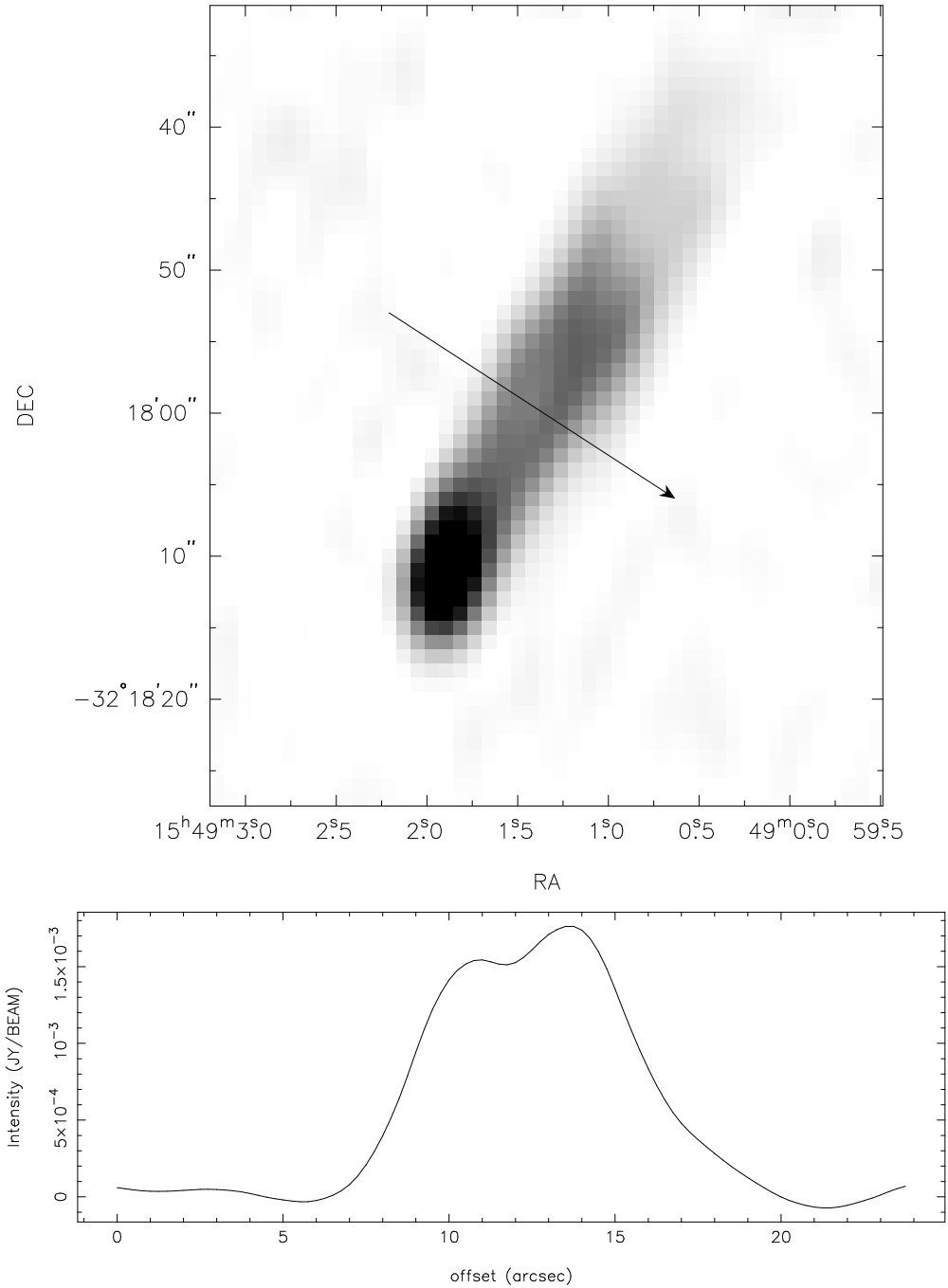}
\caption{Shows the S2 component at 12 cm with a 
beam of FWHM $5\farcs1 \times 2\farcs4$  at a P.A. of $-0\fdg8$.  
The upper panel shows the component using 
linear gray scales in the range 0--3 mJy beam$^{-1}$; the lower panel shows 
a slice profile across the inner lobe taken along the line indicated
in the upper panel. \label{fig7}
}
\end{figure}

\clearpage

\begin{figure}
\epsscale{1.0}
\plotone{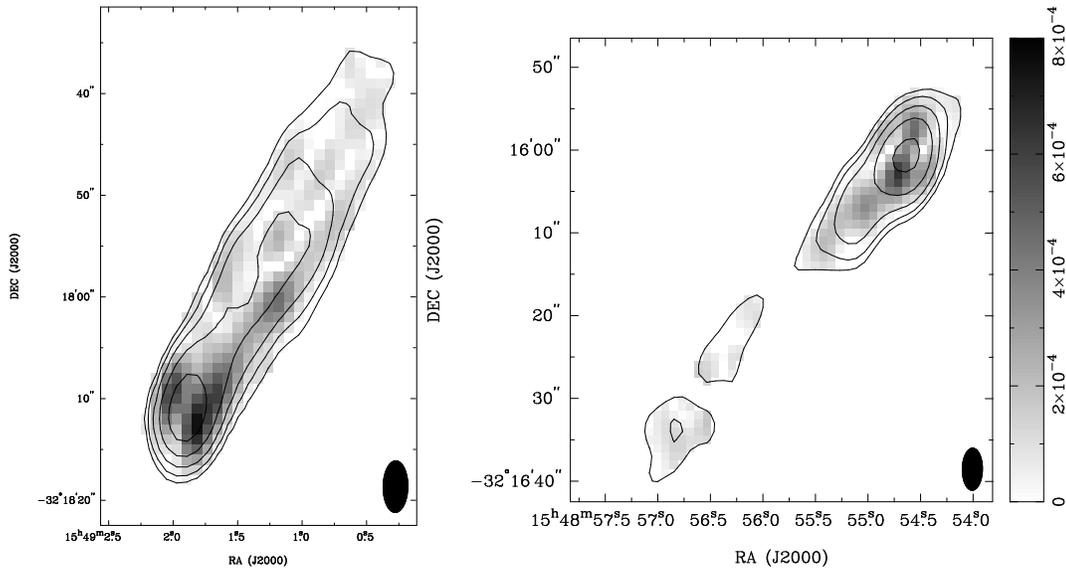}
\caption{Shows the distribution of the 12 cm polarized intensity  
over the inner lobes with a beam of FWHM
$5\farcs1 \times 2\farcs4$  at a P.A. of
$-0\fdg8$.  The polarized intensity is shown using linear gray scales
superposed on contours of the total intensity; contours are at 0.1, 0.2, 
0.4, 0.8 and 1.6 mJy beam$^{-1}$.  S2 is shown in the panel on the left,
N2 is in the panel on the right. \label{fig8}
}
\end{figure}

\clearpage

\begin{figure}
\epsscale{1.0}
\plotone{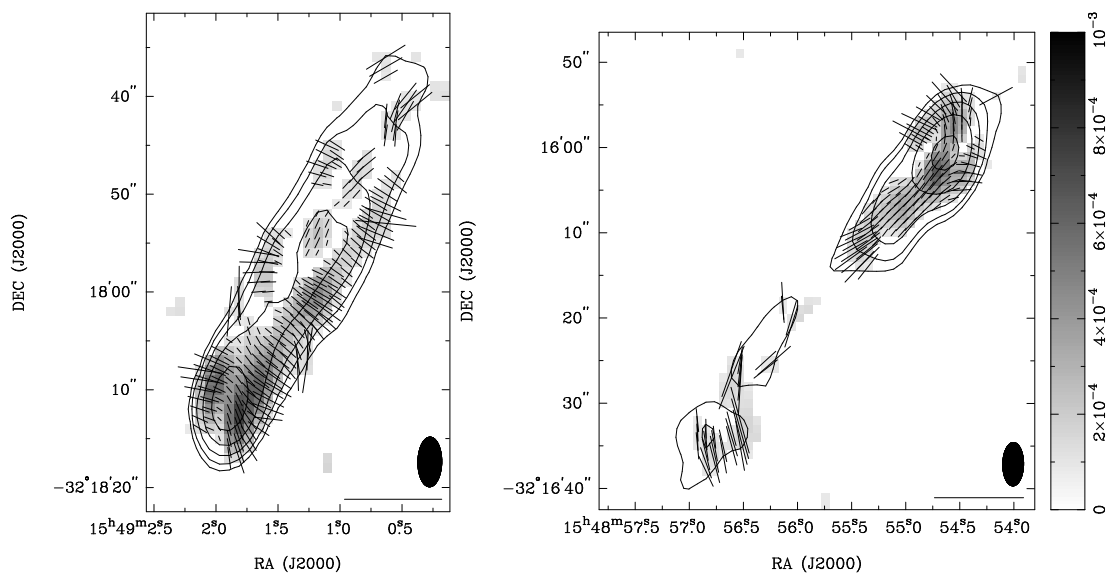}
\caption{Shows the polarization over the inner lobes of PKS~B1545$-$321: 
S2 is on the
left panel, N2 is shown on the right.  The images have been made with a beam of
FWHM $5\farcs1 \times 2\farcs4$  at a P.A. of
$-0\fdg8$. Bars show the E-field orientation, corrected for Faraday
rotation, with lengths proportional to the fractional 
polarization at 12 cm; the vectors
shown in the bottom right corners correspond to 100\% polarization. 
12-cm polarized intensity is shown using gray scales in the
range 0--0.8~mJy~beam$^{-1}$.  
Contours of the 12 cm total intensity are overlaid; contours are at 0.1, 0.2, 
0.4, 0.8 and 1.6 mJy beam$^{-1}$. \label{fig9}
	}
\end{figure}

\clearpage

\begin{figure}
\epsscale{1.0}
\plotone{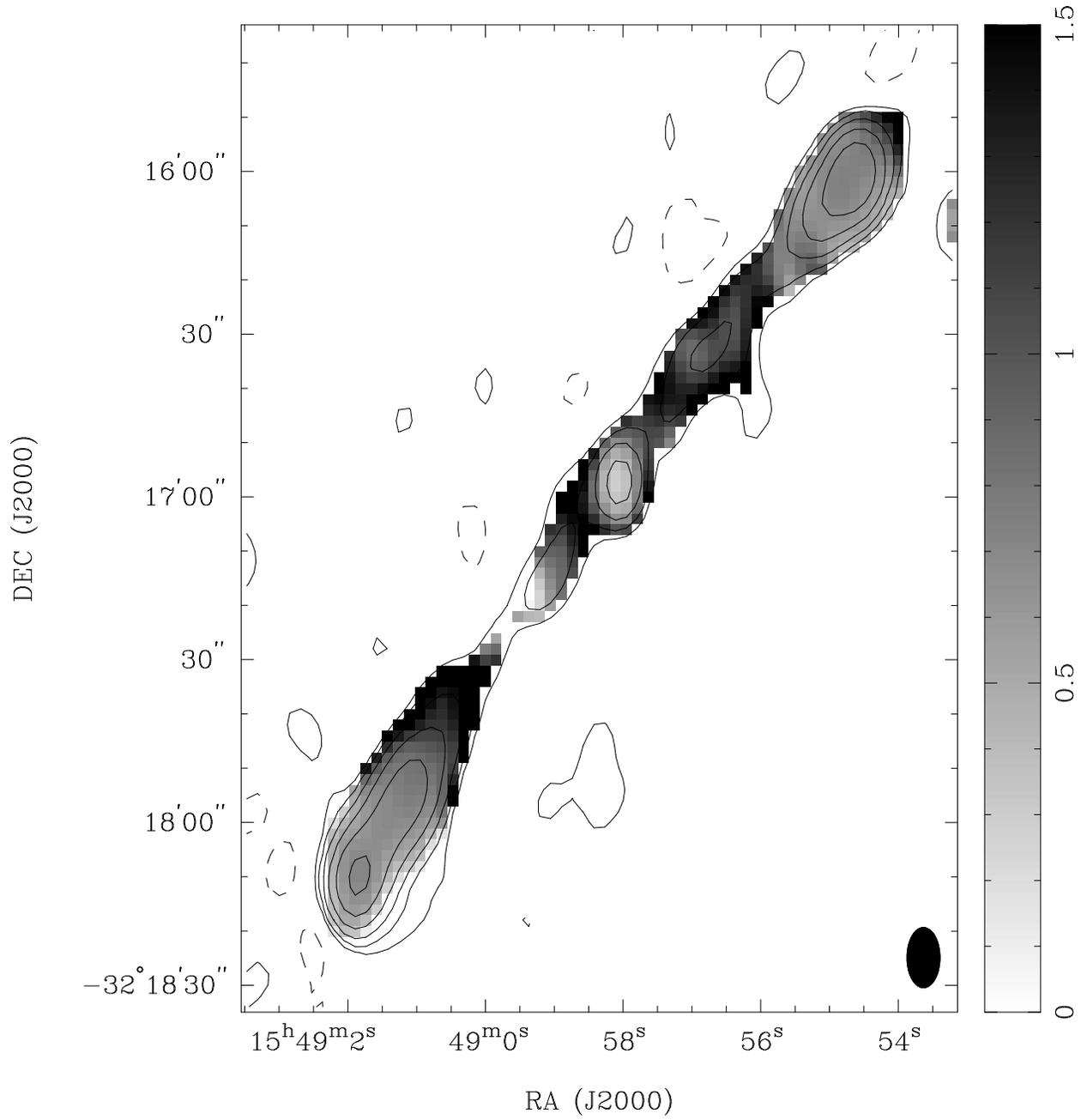}
\caption{Spectral index distribution over the inner lobes as computed 
from images at 12 and 22~cm made with beams of 
FWHM $11\arcsec \times 6\arcsec$ 
at a P.A. of 0$\degr$.  The spectral index $\alpha$ is shown using 
gray scales in the range 0--1.5. Contours of the
12 cm total intensity, at 0.3~mJy~beam$^{-1}$ x ($-1$, 
1, 2, 4, 8, 16 and 32), are overlaid. \label{fig10}
	}
\end{figure}

\clearpage

\begin{figure}
\epsscale{1.0}
\plotone{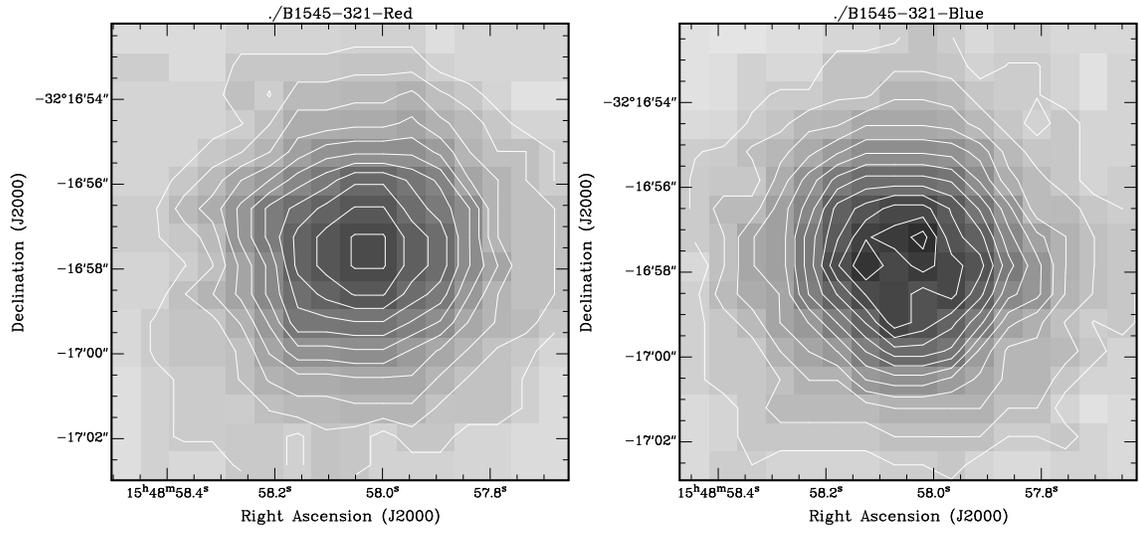}
\caption{The host galaxy. The left panel shows the SCOS digitization of the
red UK Schmidt survey image and the right
panel shows the blue image. \label{fig11}
	}
\end{figure}

\begin{figure}
\epsscale{1.0}
\plotone{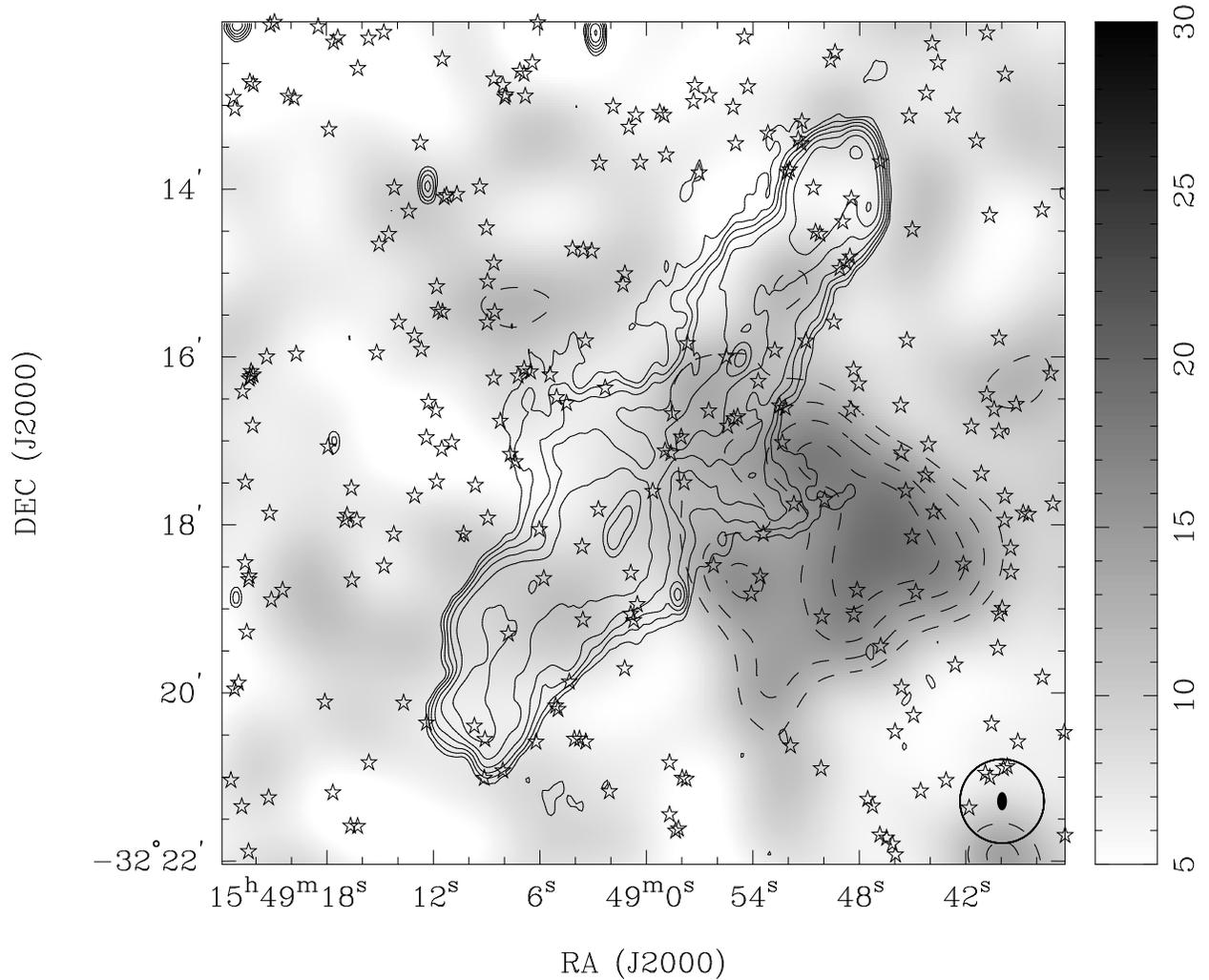}
\caption{ Radio contours overlaid on 
an archival ROSAT PSPC X-ray image of the field.  The radio contours
of the 22-cm image, made with a beam of FWHM $11\farcs6 \times 5\farcs7$ 
at a P.A. of $1\fdg6$, are at 0.2 mJy beam$^{-1}$ x (1, 2, 4, 8, 16, 32
and 64).  The PSPC pixel counts, smoothed with a $1\arcmin$ FWHM gaussian,
are shown using gray scales and dashed contours at 90\%, 80\%, 70\% and 60\% of
the peak.  Positions of galaxies in the field, within $\pm3$ mag. of the 
optical ID of the radio source, are shown with star symbols. \label{fig12}
}
\end{figure}

\end{document}